\documentclass[twocolumn,amsmath,amssymb,nofootinbib]{revtex4-2}
\usepackage{dcolumn}

\usepackage{hyperref}
\usepackage{graphicx}
\usepackage{bm}
\usepackage{xcolor}

\begin{document}
	
	\title{A Hall viscosity for skyrmion via magnon interaction}
	
	\author{Bom Soo Kim}
	\affiliation{%
		Department of Mathematics and Physics, University of Wisconsin-Parkside,  Kenosha WI 53141,
		USA
	}%
	
	\date{\today}

	\vspace{0.5in}
	\begin{abstract}
		We identify a Hall viscosity term directly from the Dzyaloshinskii-Moriya interaction (DMI), that breaks parity symmetry, in the skyrmion motion of insulating magnets by time-averaging the magnon contribution to all orders. The viscosity term is proportional to the skyrmion charge. Skyrmion Hall angle shows significant dependence on the skyrmion shape and size, the ratio of exchange over DMI parameters, while roughly independent of the Gilbert damping parameter. The Hall angles have the same magnitude for opposite skyrmion charges. We speculate a velocity-dependent Hall viscosity contribution to seek asymmetric Hall angles for the opposite charges.       
	\end{abstract}
	
	\maketitle
	\noindent

\section{Introduction} 

Hall viscosity \cite{Avron:1995} is a fundamental and universal transport coefficient in the absence of parity symmetry. Hydrodynamics has confirmed its existence \cite{Jensen:2011xb,Bhattacharya:2011tra}, and there have been a plethora of investigations in the context of quantum Hall systems \cite{Read:2008rn,Hoyos:2011ez,Bradlyn:2012ea,Hoyos:2014lla,Hoyos:2015yna}. Recently it is introduced to skyrmion physics using topological Ward identities in \cite{Kim:2015qsa,Kim:2019vxt} and various different ways to verify its existence have been proposed \cite{Kim:2020piv,Kim:2023zlm,Kim:2023BB}. These topics of Hall viscosity in magnetic skyrmions have been also collected in a monograph \cite{KimBook}. Moreover, Hall effects haven been also observed in charge neutral spin objects such as skyrmions \cite{Jiang2017,Litzius2017mmm} and magnons  \cite{MagnonHall1} and even for spin charge neutral phonons \cite{PhononHall1}.  

Here we investigate the Hall viscosity contributions in the skyrmion motion by including the skyrmion magnon interactions in the context of insulating magnets. We note there exist closely related works in literature on Landau-Lifshitz-Gilbert (LLG) equation in this context \cite{MagnonCourseGraining1,MagnonCourseGraining2,MagnonCourseGraining3,MagnonCourseGraining4,MagnonCourseGraining5,MagnonCourseGraining6}. We generalize the results by including the magnon density to all orders. This work can be viewed as a part of bigger program of investigating Hall viscosity in skyrmion motion 
by extending the previous works on conducting magnets and also insulating magnets without magnon contributions \cite{Kim:2015qsa,Kim:2019vxt,KimBook}. We provide derivations fully in appendices.   

At the outset, we clarify a few subtle technical details. To study the interaction between skyrmion and magnon, we use the decomposition of a magnetization vector as
\begin{align}  \label{MagnetizationDecomposition}
\vec m &= (1 - m_f^2)^{1/2} \vec m_s +  \vec m_f  \;. 
\end{align}	 
with the slow mode $\vec m_s$ and the fast mode $\vec m_f$, which is typically much smaller than the slow mode $|\vec m_f| \ll |\vec m_s|$. The factor $  (1 - m_f^2)^{1/2} $ is to ensure that both vectors $\vec m $ and $ \vec m_s $ have the unit norms as $|\vec m^2| =1 $ and $ |\vec m_s^2| =1 $. Here we consider the case when $\vec m_f$ has a periodic time dependence, which allows us to average over the fast mode in time (temporal coarse graining). This can be further justified as we are interested in steady-state motion of topological defects such as skyrmions and domain walls. Furthermore, we do not expand the square root factor aiming to get the full results without approximation, which is theoretically appealing. 

We also note that the slow mode is also slowly varying in space compared to the fast mode, $|\vec \nabla_\alpha  \vec m_s| \ll |\vec \nabla_\alpha \vec m_f|$ where $\alpha =1,2$ are the space indices. It is tempting to think that the terms combined with $\vec m_f$ and $\vec \nabla_\alpha \vec m_s$ are smaller than those combined with $\vec m_s$ and $\vec \nabla_\alpha \vec m_f$. For example, $(\vec \nabla_\alpha \vec m_s) \cdot \vec m_f  \ll  \vec m_s \cdot (\vec \nabla_\alpha \vec m_f)$. This is not the case as the two modes, $ \vec m_s $ and $ \vec m_f $, are perpendicular to each other and constrained to satisfy $\vec m_s \cdot \vec m_f = 0 $. Thus, they satisfy 
\begin{align}  \label{DerivativeISlowFastModes}
(\vec \nabla_\alpha \vec m_s) \cdot \vec m_f  = - \vec m_s \cdot (\vec \nabla_\alpha \vec m_f) \;, 
\end{align} 
where $\cdot$ acts on the two magnetization vectors and $\vec \nabla_\alpha$ is gradient operator. Thus, we carefully keep all the terms when spatial derivatives are involved. 

Also, we do not consider the spatial averaging for the fast modes as it is not clear how to take spatial averages consistently when involved with spatial derivatives. For example, we have the identity 
\begin{align}
\vec m_f (\vec \nabla_\alpha \vec m_s \cdot \vec m_f ) = - \vec m_f ( \vec m_s \cdot \vec \nabla_\alpha \vec m_f) \;.
\end{align} 
The term in the left side gives a finite result while the right one vanishes if we average them over the spatial directions, say for the fast mode $\vec m_f \propto \sin (k_x x) \sin (k_y y)$ and the domains $ 0 \leq x, y < 2\pi/k$. 

In main sections, we construct the LLG equation and the Thiele equation for skyrmion by using the decomposition \eqref{MagnetizationDecomposition} and the temporal time averaging. Then we evaluate them with a continuous skyrmion model along with a circular symmetric magnons. The Hall viscosity term and the corresponding Hall angles are analyzed with opposite skyrmion charges. We seek to find a way for achieving asymmetric skyrmion angles before conclude.   

\section{Temporal averaging of Magnons} \label{sec:TemporalTimeAverage}

There have been numerous results of separating the slow mode and the fast mode \eqref{MagnetizationDecomposition}for the Landau-Lifshitz-Gilbert (LLG) equation  
\begin{align} \label{LLGeq}
	\dot {\vec m} = -\gamma \vec m \times \vec H_{\text{eff}} + \alpha \vec m \times \dot {\vec m} \;, 
\end{align} 	
in the context of spins or magnetizations, e.g.,   \cite{MagnonCourseGraining1,MagnonCourseGraining2,MagnonCourseGraining3,MagnonCourseGraining4,MagnonCourseGraining5,MagnonCourseGraining6}. Here $\vec H_{\text{eff}} = -\delta \mathcal H/\delta \vec m $ with Hamiltonian $\mathcal H$ and only the precession term is captured by $\mathcal H$.

\newpage 

We study the time average of the fast mode with a periodic time dependence, after rewriting the magnetization with \eqref{MagnetizationDecomposition}. Starting from the terms with the time derivatives, we get  
\begin{align} \label{LLGEQFirstTerm}
	\dot {\vec m} &= (1 - m_f^2)^{1/2} \dot{\vec m}_s + \dot{\vec m}_f - \frac{(m_f^k \dot{m}_f^k)}{ (1 - m_f^2)^{1/2} } \vec m_s  \nonumber \\
	&\rightarrow    (1 - \langle m_f^2 \rangle_T )^{1/2}  \dot{\vec m}_s \;, 
\end{align}
where the second term, linear in $ \vec m_f$, vanishes for the temporal averaging over a period. The third term, while quadratic, also vanishes as we will see later with a circular symmetric form of the fast mode. Similarly, the damping term, the last term in \eqref{LLGeq}, after dropping the odd power of the fast mode, reduces 
\begin{align}
	\vec m \times \dot {\vec m} 
	&\rightarrow  (1 - \langle m_f^2 \rangle_T) \vec m_s \times \dot{\vec m}_s + \langle \vec m_f \times \dot{\vec m}_f \rangle_T  \;. 
\end{align}
Note the last term $ \langle \vec m_f \times \dot{\vec m}_f \rangle_T $ survives with the time dependence we choose and points to the direction of the slow mode \cite{footnote1}. 

The second term in \eqref{LLGeq} can be also evaluated with the  Hamiltonian $ \mathcal H =  \int dV \Big( \frac{J}{2}~ (\vec \nabla \vec m)^2 + D ~\vec m \cdot (\vec \nabla \times \vec m) - \vec H \cdot \vec m\Big) $. The form $\vec H_{\text{eff}} = -\delta \mathcal H/\delta \vec m $  and detailed computations can be found in appendix \S \ref{app:EvaluatingLLG}. The resulting LLG equation after temporal average reads

\begin{widetext}
	
\begin{align} \label{LLGeqWithMagnonContributions}
\dot {\vec m}_s =& -\gamma \vec m_s \times \vec H_{\text{eff}}^{0s} +  (1 - \langle m_f^2 \rangle)^{1/2} \cdot \alpha \vec m_s \times \dot {\vec m}_s  + \langle \vec m_f \times \dot{\vec m}_f \rangle_T  \\
& - \frac{\gamma }{(1 - \langle m_f^2 \rangle)^{1/2} } \Big\{  J ~\langle \vec m_f \times \nabla_\alpha^2 \vec m_f \rangle_T   
- J ~ \langle m_f^i \nabla_\alpha m_f^i \rangle_T [\vec m_s \times  \nabla_\alpha \vec m_s]    \nonumber  \\
&\qquad\qquad\qquad\qquad - 2D ~ \langle \vec m_f \times ( \nabla \times \vec m_f) \rangle_T  
+ 2D ~\vec m_s \times [( \vec \nabla \langle m_f^2 \rangle_T) \times \vec m_s   \Big\}
\;, \nonumber
\end{align} 	 
where $\vec H_{\text{eff}}^{0s} = (1 - \langle m_f^2 \rangle)^{1/2} \{J  (\nabla^2 \vec m_s  ) 
-  2D (\vec \nabla \times  \vec m_s) \} + \vec H  $.  
Similar results were previously reported  \cite{MagnonCourseGraining1,MagnonCourseGraining2,MagnonCourseGraining3,MagnonCourseGraining4,MagnonCourseGraining5,MagnonCourseGraining6}. We include all the magnon contributions with only its temporal average. Detailed derivations can be found in \S \ref{app:EvaluatingLLG}.

Thiele equation describes the linear motion of the slow mode center, $\vec m_s (r_\alpha (t)) $, where $\alpha, \beta = 1,2 $ for 2 dimensional space. Thus, $ \dot{\vec m}_s  =  \dot{r}_\beta  \nabla_\beta \vec m_s = v_\beta \nabla_\beta \vec m_s $. By contracting \eqref{LLGeqWithMagnonContributions} with $(\vec m_s \times \nabla_\beta \vec m_s ) ~\cdot $ ($\cdot$ acts on $\vec m$), we get 
\begin{align} \label{ThieleEqMagnonAveraged}
 	0 =& - \int dV  \mathcal G_{\alpha\beta} v_\beta  +  \alpha  \int dV (1 - \langle m_f^2 \rangle)^{1/2} \mathcal D_{\alpha\beta} v_\beta \\
 	& -\gamma \int dV \left\{ (\nabla_\alpha \vec m_s) \cdot \vec H +  (1-m_f^2)^{1/2} \left[ (J)(\nabla_\alpha \vec m_s \cdot \nabla^2 \vec m_s)  - (2D) (\nabla_\alpha \vec m_s \cdot \vec \nabla \times \vec m_s) \right] \right\} \nonumber   \\
 	&-\gamma (J) \int dV \Big\{ - \frac{ [ (\vec m_f \cdot \nabla_\beta \vec m_f)] [(\nabla_\alpha \vec m_s) \cdot (\nabla_\beta \vec m_s)] }{(1 - \langle m_f^2 \rangle)^{1/2} } -  \frac{ [ \vec m_s \cdot \nabla^2 \vec m_f] [\vec m_f \cdot  \nabla_\alpha \vec m_s]  }{(1 - \langle m_f^2 \rangle)^{1/2} }  \Big\} \nonumber \\
 	&+ \gamma (2D) \int dV \Big\{ \frac{ [\vec m_s \cdot (\vec \nabla \times \vec m_f)]  (\vec m_s \cdot \nabla_\alpha \vec m_f) }{(1 - \langle m_f^2 \rangle)^{1/2} }  -  \frac{ (m_f^i \vec \nabla m_f^i) \cdot [ \vec m_s \times  \nabla_\alpha \vec m_s]  }{(1 - \langle m_f^2 \rangle)^{1/2} }  \Big\} 
 	\;, \nonumber
\end{align}
\vspace{-0.12in}

\end{widetext}

\noindent 
where the integral is over a single skyrmion unit. Detailed derivation of \eqref{ThieleEqMagnonAveraged} can be found in \S \ref{app:ExampleThieleEq}. Here, $\mathcal G_{\alpha\beta}=\vec m_s \cdot (\nabla_\alpha \vec m_s \times \nabla_\beta \vec m_s)$ is directly related to the skyrmion charge, while $\mathcal D_{\alpha\beta}=(\nabla_\alpha \vec m_s) \cdot (\nabla_\beta \vec m_s)$ is a drag term. While the last line contains 2 derivatives, one for slow mode and one fast, the derivative for the fast mode can be switched to the slow mode using \eqref{DerivativeISlowFastModes}. It turns out that the terms are proportional to the skyrmion charge. The integral contains the thickness of the ferromagnetic thin film, that is assumed to cancel out among the terms.

\section{Continuous Skyrmion model}  	

We evaluate the Thiele equation \eqref{ThieleEqMagnonAveraged} using the slow mode that satisfies $ \vec m_s^2 =1$ as 
 \begin{align} \label{LocalCoordinatesI}
	\vec m_s = &\hat e_3 = \sin \theta \cos \phi ~\hat x +  \sin \theta \sin \phi ~\hat y + \cos \theta ~\hat z  \;,
\end{align}
where $\theta$ and $\phi$ depends on the coordinates $(\rho, \varphi)$ or $(x=\rho \cos \varphi,y=\rho \sin \varphi)$ as $\theta (\rho)$ and $\phi (\varphi)$. The fast mode, transverse to the slow mode, can be described as 
 \begin{align} \label{MagnonWaveFunctionCircular}
 	\vec m_f = & \psi (\rho, \varphi) e^{-i\omega t} \hat e_+  +  \psi^* (\rho, \varphi) e^{i\omega t} \hat e_-  \;, 
 \end{align}
where $\psi (\vec r) = \psi_R (\vec r) + i \psi_I (\vec r) $ and $\hat e_\pm = (\hat e_1 \pm i \hat e_2)/2 $ with $\hat e_1 = \cos \theta \cos \phi ~\hat x +  \cos \theta \sin \phi ~\hat y - \sin \theta ~\hat z $, $\hat e_2 = - \sin \phi ~\hat x +  \cos \phi ~\hat y$. Here we focus on the `circular symmetric' magnon with respect to slow mode $\vec m_s$. Thus, $ \vec m_f = \psi_1 \hat e_1 + \psi_2 \hat e_2$ with $ \psi_1=  \psi_R (\vec r) \cos (\omega t) + \psi_I (\vec r) \sin (\omega t)  $ and $  \psi_2 = \psi_R (\vec r) \sin (\omega t) - \psi_I (\vec r) \cos (\omega t)  $. 
Then, the temporal time average gives $\langle \psi_1^2 \rangle_T = \langle \psi_2^2 \rangle_T = \langle \psi^2 \rangle_T /2$. A temperature gradient is applied along $x$ direction and thus $ \nabla_x \langle \psi^2 \rangle_T \neq 0$ (and assumed to be constant over the unit cell of the skyrmion) and $ \nabla_y \langle \psi^2 \rangle_T = 0 $. 

We consider a single skyrmion that has the angular profile with a finite volume, $0 \leq \rho < 2P, 0 \leq \varphi < 2\pi$. 
 \begin{align} \label{SmoothSkyrmionModel}
\theta(\rho) = \left\{ \begin{matrix}
\pi \;, &  & 0 < \rho  < P -  \frac{\omega_{\text{DW}}}{2}  \; \\ \vspace{-0.1in}
& & \\
\frac{\pi}{2} - \frac{\rho - P }{\omega_{\text{DW}}} \pi \;,   & & P - \frac{\omega_{\text{DW}}}{2}  \leq \rho  \leq P +  \frac{\omega_{\text{DW}}}{2} \\  \vspace{-0.1in}
& & \\
0 \;, &  & P +    \frac{\omega_{\text{DW}}}{2} < \rho < 2P   \;
\end{matrix}
\right.
\end{align}
and identify $ \phi = \varphi$. The second line of \eqref{SmoothSkyrmionModel} describes the domain wall (DW) in the middle range of the skyrmion. Then the Thiele equation reads 
\begin{align} \label{ThieleEqMagnonCoordinates}
0 &=  4\pi \epsilon_{\alpha\beta} v_\beta + \alpha (1-\langle \psi^2 \rangle)_T )^{1/2} \delta_{\alpha\beta} v_\beta ~\pi \Big[  \frac{\pi^2  P}{\omega_{\text{DW}}} +C_{\omega_{\text{DW}}} \Big]  \nonumber 
\\
&- \hat x ~(J \gamma )~ \frac{\pi}{2} \Big[  \log \big[ \frac{2P}{\rho_{min}} \big] - C_{\omega_{\text{DW}}} \Big] \cdot \frac{ \nabla_x \langle \psi^2 \rangle_T}{(1 - \langle \psi^2 \rangle_T)^{1/2}} \nonumber \\
&+ \hat y ~(2D \gamma)~ \Big[  \frac{\pi^2 P}{4} \Big] \cdot \frac{  \nabla_x \langle \psi^2 \rangle_T}{(1 - \langle \psi^2 \rangle_T)^{1/2}} \;.  
\end{align}
The first line of \eqref{ThieleEqMagnonCoordinates} contains the well known skyrmion Hall effect term, transverse to the direction of motion, and the longitudinal drag term, that depends on the details of the skyrmion structure and also the magnon density. The skyrmion with skinny DW region has bigger drag effect compared to the skyrmion with fat DW region. Here, $C_{\omega_{\text{DW}}}  = \int_{P-\omega_{\text{DW}}/2}^{P+\omega_{\text{DW}}/2} (\sin^2 \theta /\rho) d\rho$ only depends on the ratio $P/ \omega_{\text{DW}} $ and is much smaller than the other contribution as we see below. Analytic result of $C_{\omega_{\text{DW}}}$ is listed in \eqref{CDIntegral} in the appendix \S \ref{app:ExampleThieleEq}. We also note that the second line of \eqref{ThieleEqMagnonAveraged} vanishes as they are circularly symmetric except the gradient $\nabla_\alpha$, which picks a direction along $\alpha = x, y$ with an integrating factor $\sin \varphi$ or $\cos \varphi$ for $\varphi$ integral.  

This analytic result \eqref{ThieleEqMagnonCoordinates} also contains the force density contributions that is parallel and perpendicular to the gradient of magnon density. The latter is the same as the direction of applied temperature gradient. The second line of \eqref{ThieleEqMagnonAveraged} is the combination of two terms that are involved with $J$ with $ \log [ {2P}/{\rho_{min}} ] - C_{\omega_{\text{DW}}}>0 $. Detailed computations are provided in \S \ref{app:ExampleThieleEq}. Note that the $- \nabla_x \langle \psi^2 \rangle$ tells that the direction of force is opposite to the gradient of magnon density, the direction of magnon motion. We also note that it depends on the details of skyrmion structure, the area $2\pi P {\omega_{\text{DW}}}$ of domain wall region of the skyrmion. 

The last term in \eqref{ThieleEqMagnonCoordinates} is the sought contribution that is transverse to the direction of the temperature gradient. The magnitude is directly proportional to $2\pi P$, the circumference of the middle of circular skyrmion, independent of the thickness of the DW inside the skyrmion. We advertise that this is an explicit example of Hall viscosity contribution that is originated from the term $ - D \vec m \cdot (\vec \nabla \times \vec m)$ in Hamiltonian which breaks the parity symmetry. Note that the term is generated through the interaction between the skyrmion and magnon that is assumed to have effects along the direction of temperature gradient, $ \nabla_x \langle \psi^2 \rangle$. 

Now solving \eqref{ThieleEqMagnonCoordinates} gives 
\begin{align} \label{SolutionThieleEq}
v_x &= \frac{ \tilde \alpha A + 4\pi B}{(4\pi)^2 + \tilde \alpha^2} \frac{\nabla_x \langle \psi^2 \rangle_T }{(1 - \langle \psi^2 \rangle_T)^{1/2}}  \;, \nonumber \\
v_y &=  \frac{ 4\pi A -  \tilde \alpha B }{(4\pi)^2 + \tilde \alpha^2} ~ \frac{\nabla_x \langle \psi^2 \rangle_T }{(1 - \langle \psi^2 \rangle_T)^{1/2}}  \;, 
\end{align}
where 
$
\tilde \alpha = \alpha (1-\langle \psi^2 \rangle)_T )^{1/2} \pi  \cdot C_\alpha 
$ with $ C_\alpha =   \frac{\pi^2  P}{\omega_{\text{DW}}} +C_{\omega_{\text{DW}}}  $,  
$
A 
= (\pi/2) \gamma J  \cdot  C_A$ with $C_A = \log \big[ \frac{2P}{\rho_{min}} \big]  - C_{\omega_{\text{DW}}}
$ 
and 
$
B = (\pi^2/2)  \gamma D P.
$
$C_\alpha$ and $ C_A$ only depends on the details of the skyrmion profile. 
We can see that the terms $\tilde \alpha, A, B$ are sensitive to the details of the skyrmion structure. In particular $\tilde \alpha$ only depends on the ratio $P/\omega_{WD}$. On the other hand, $A$ depends on the size and the cutoff of lower radial integral domain of the skyrmion, while $B$ only depends on the size. 
From the analytic form of the first equation in \eqref{SolutionThieleEq}, the longitudinal velocity along the temperature gradient behaves $v_x \propto \nabla_x \langle \psi^2 \rangle \propto \nabla_x T  $ as its coefficient is positive and magnon density is proportional to temperature. Thus the skyrmion \cite{MagnonCourseGraining2,MagnonCourseGraining3,MagnonCourseGraining4} and Domain Wall \cite{MagnonCourseGraining6}  move toward the hotter region 
. 

\section{Hall angles} 

The Hall angle can be obtained from \eqref{SolutionThieleEq} as $\tan \theta_H = v_y/v_x =  {(4\pi A - \tilde \alpha B) }/{ (\tilde \alpha A+ 4\pi B )}  $, where the magnon contributions and details of driving force cancel out. Here we choose $\langle \psi^2 \rangle_T \ll 1$. Instead of estimating the Hall angle with a set of particular parameter values, we rewrite the Hall angle in terms of the parameters $J/D $, $\alpha$ and the parameters depending on skyrmion structure, $P, C_\alpha, C_A$.  
\begin{align} \label{HallAngleDetail}
	\tan \theta_H = \frac{4 C_A(J/D) - \pi \alpha C_\alpha P  }{ \alpha C_\alpha C_A (J/D)+ 4 \pi P }   \;. 
\end{align} 
We note that the Hall angle depends on $J$ and $D$ only through the ratio $J/D$. For a reasonable range of parameters, $C_\alpha \sim \mathcal O(10)$ and $C_A \sim \mathcal O(1)$. For example, $C_\alpha =  23.71 $, $C_A = 4.39$ for $P= 50 ~nm, \omega_{WD} = 21~ nm $ and $ \rho_{min} =1 ~nm$. 

\begin{figure}[t!]
	\begin{center}
		\includegraphics[width=0.45\textwidth]{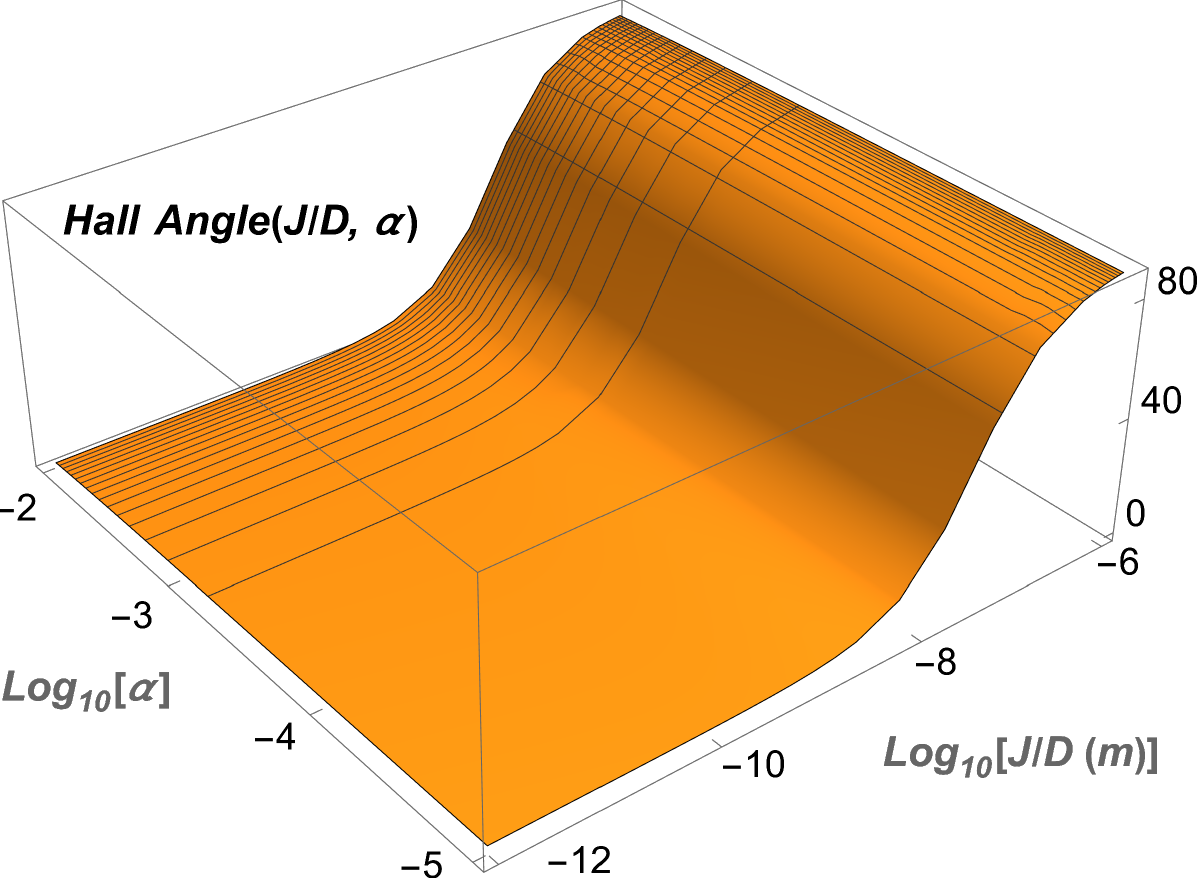}
		\caption{\footnotesize\small  Hall angle with Log-Log scales for $ (J/D, \alpha)$ \vspace{-0.2in} 
		}
		\label{fig:HallSkyrmionModel}
	\end{center}
\end{figure}

We consider the Hall viscosity dependence on the three parameters with ranges as $ 10^{-12} m < J/D < 10^{-6} m, P \sim 10^{-8} m, 10^{-5} < \alpha < 10^{-2}$ for a continuum skyrmion model. Note that the denominator in \eqref{HallAngleDetail} is roughly order of $10^{-7} - 10^{-6} ~m $ due to the fixed value $4\pi P \sim 10^{-7}~ m$ 
\cite{footnote2}. Thus, the Hall angle is large for a large portion of the parameter space (note that the plot has a log scale). It changes steeply when $ J/D \sim P$ and $\tan \theta_H \sim { C_A(J/D) }/{  (\pi P) } $. The Hall angle (in degree) is depicted in Fig. \ref{fig:HallSkyrmionModel} with the values of $J/D$ and $\alpha$ in log-log scales with base $10$, e.g. $ Log[\alpha] = -2 $ for $\alpha = 10^{-2}$. Note that the Hall angle strongly depends on the combination $J/D$, increasing steeply around $J/D \sim 10^{-7} ~m $. 

The result tells that the skyrmion size has a significant role in Hall angle, especially when $ \alpha \cdot (J/D) $ is smaller than the skyrmion size $P$, while the angle is almost independent of $\alpha$. With a smaller size skyrmion, $P= 10 ~nm, \omega_{WD} = 5~ nm $ and $ \rho_{min} =1 ~nm$, the Hall angle increases steeply around $J/D \sim 10^{-8} ~m $, an order of smaller value of $J/D$.  

When the skyrmion has an opposite topological charge, the Thiele equation changes the signs of the transverse terms, the first term with $\epsilon_{\alpha\beta}$ and the last term that is proportional to $D$. Thus the velocities \eqref{SolutionThieleEq} are the same with a relative sign for $v_y$, and the corresponding Hall angle \eqref{HallAngleDetail} also changes sign. This is expected as both the terms, the Magnus force and the DM term, are related to the magnetic properties, meaning to be involved with a charged transport rather than neutral one. 

Now, it is interesting to see whether the broken parity symmetry in a charged sector also breaks the parity in a neutral sector. Rephrase differently, is there a neutral Hall viscosity effect or neutral momentum transport (in addition to the momentum driven by charged dynamics) even in this case at hand, either directly or through some interactions between the charged and neutral sectors? We discuss this in the following section.

\section{Asymmetric skyrmion Hall angle? } 

We saw that the skyrmions with positive and negative charges have the same Hall angles (of course with opposite directions). It is interesting to achieve asymmetric skyrmion Hall angles. Some available experimental data appears to show discrepancies between these Hall angles \cite{Jiang2017,VanishingSkyrmionHall}. The asymmetric Hall angles provide direct ways to verify the existence of Hall viscosity as their values are different from expected ones \cite{Kim:2020piv} as well as they meet in different point compared to an expected on in the parameter space \cite{Kim:2023zlm}\cite{Kim:2023BB}. 
Moreover, it will shed a new light on how to achieve the skyrmion motion without transverse motion. See some ways achieving vanishing Hall angle by considering antiferromagnetically exchange-coupled bilayer system \cite{VanishingSkyrmionHallInBilayer}, by using two different DM interactions \cite{VanishingHallAngleWithDMIs}, or at angular momentum compensation point \cite{VanishingSkyrmionHall}.

Any moving object has a momentum regardless of its charge and is under the influence of broken parity once it is broken either at the fundamental level or through some interactions. Here we proceed to provide a way to describe the Hall viscosity that directly depends on skyrmion velocity and is independent of the skyrmion charge. This can be done by generalizing the skyrmion center collective coordinates $ r_\alpha (t)$ for $\vec m_s (r_\alpha (t)) $ as
\begin{align} \label{SkyrmionCM}
	r_\alpha &= v_\alpha t + R \epsilon_{\alpha\beta} v_\beta t \;, 
\end{align} 
which gives $\dot {\vec m}_s = (v_\alpha + R \epsilon_{\alpha\beta} v_\beta) \nabla_\alpha \vec m_s $ instead of $ \dot{\vec m}_s = v_\beta \nabla_\beta \vec m_s $. Then, the Thiele equation \eqref{ThieleEqMagnonCoordinates} is changed with two more terms by replacing $ v_\alpha $ to $ v_\alpha + R \epsilon_{\alpha\beta} v_\beta$. Solving the modified Thiele equation gives 
\begin{align}  
	v_x &= \frac{ (\tilde \alpha  \mp 4\pi R) A + (4\pi \pm \tilde \alpha R) B}{(16\pi ^2+ \tilde \alpha ^2)( 1 + R^2) } \cdot \frac{\nabla_x \langle \psi^2 \rangle_T }{(1 - \langle \psi^2 \rangle_T)^{1/2}}  \;,  \nonumber \\
	v_y &=  \frac{ (\pm 4\pi + \tilde \alpha R) A +  (\mp \tilde \alpha  + 4\pi R) B }{(16\pi ^2+ \tilde \alpha ^2)( 1 + R^2) } \cdot  \frac{\nabla_x \langle \psi^2 \rangle_T }{(1 - \langle \psi^2 \rangle_T)^{1/2}}  \;.  \nonumber 
\end{align} 
Here we include both the cases with the positive $+1$ and negative $-1$ skyrmion charges. This equation reduces to \eqref{SolutionThieleEq} for positive charge $+1$ when $R=0$.

\begin{figure}[t!]
	\begin{center}
		\includegraphics[width=0.22\textwidth]{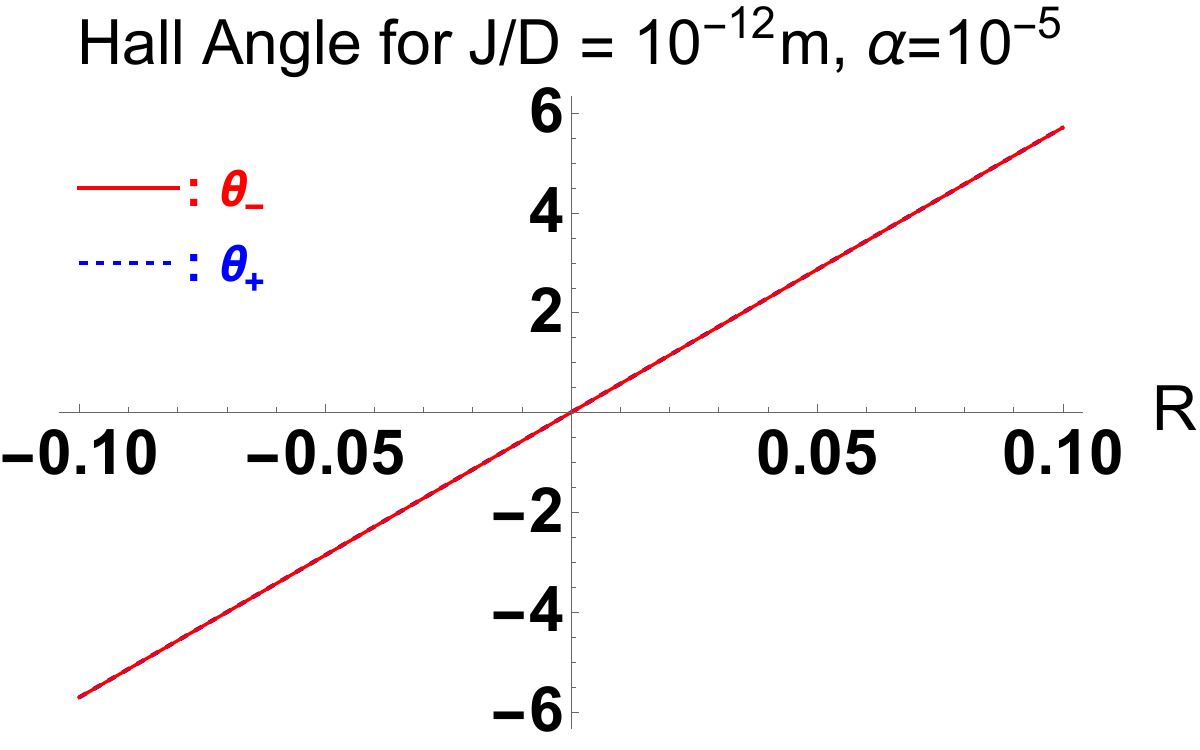} ~~ \includegraphics[width=0.22\textwidth]{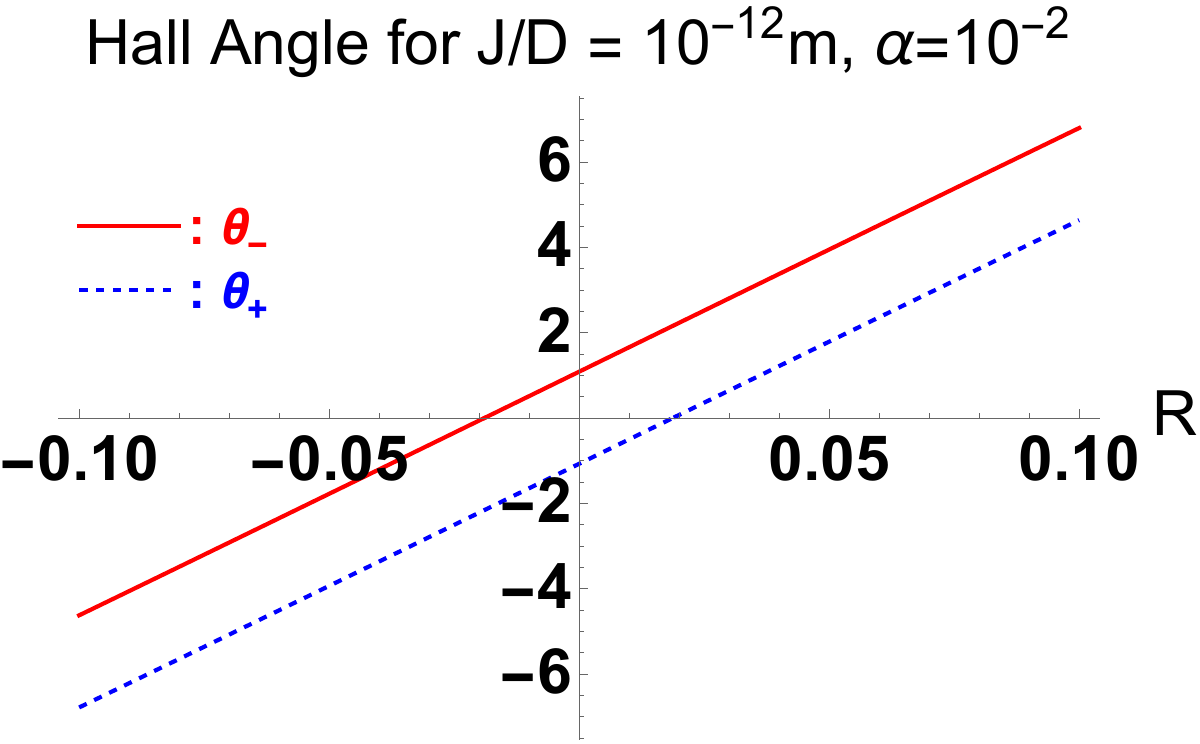} \\
		\vspace{0.15in}
		\includegraphics[width=0.22\textwidth]{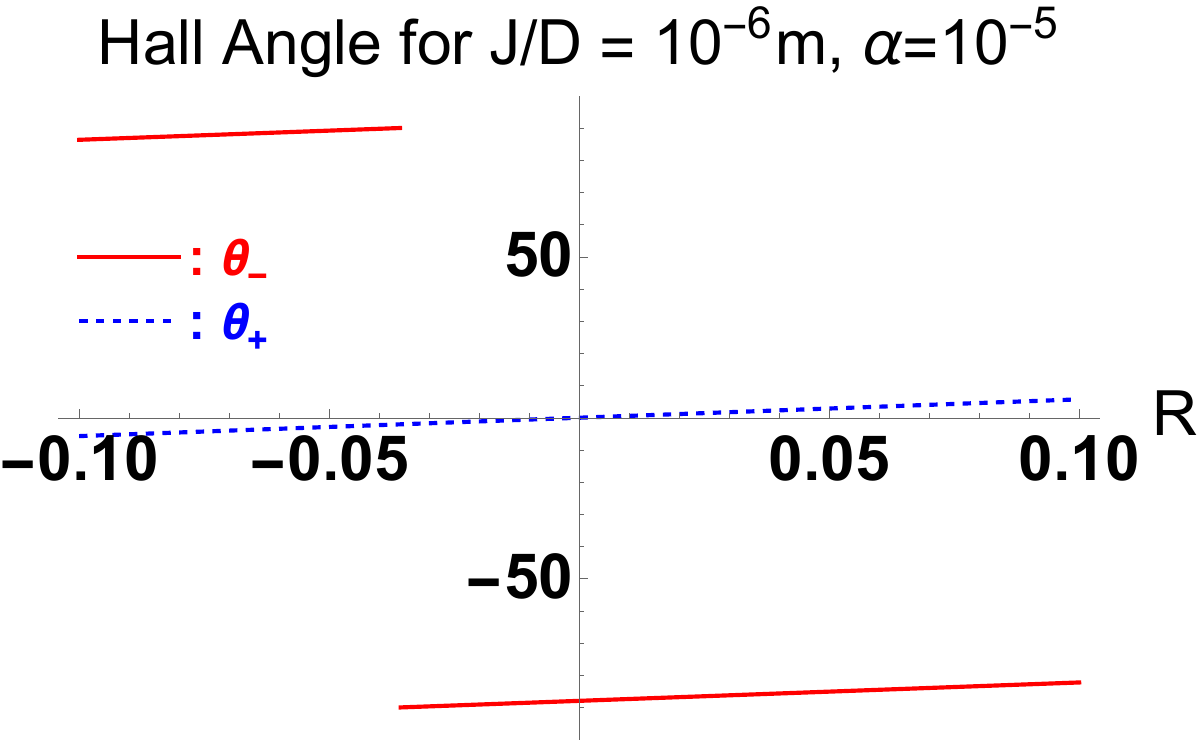} ~~
		\includegraphics[width=0.22\textwidth]{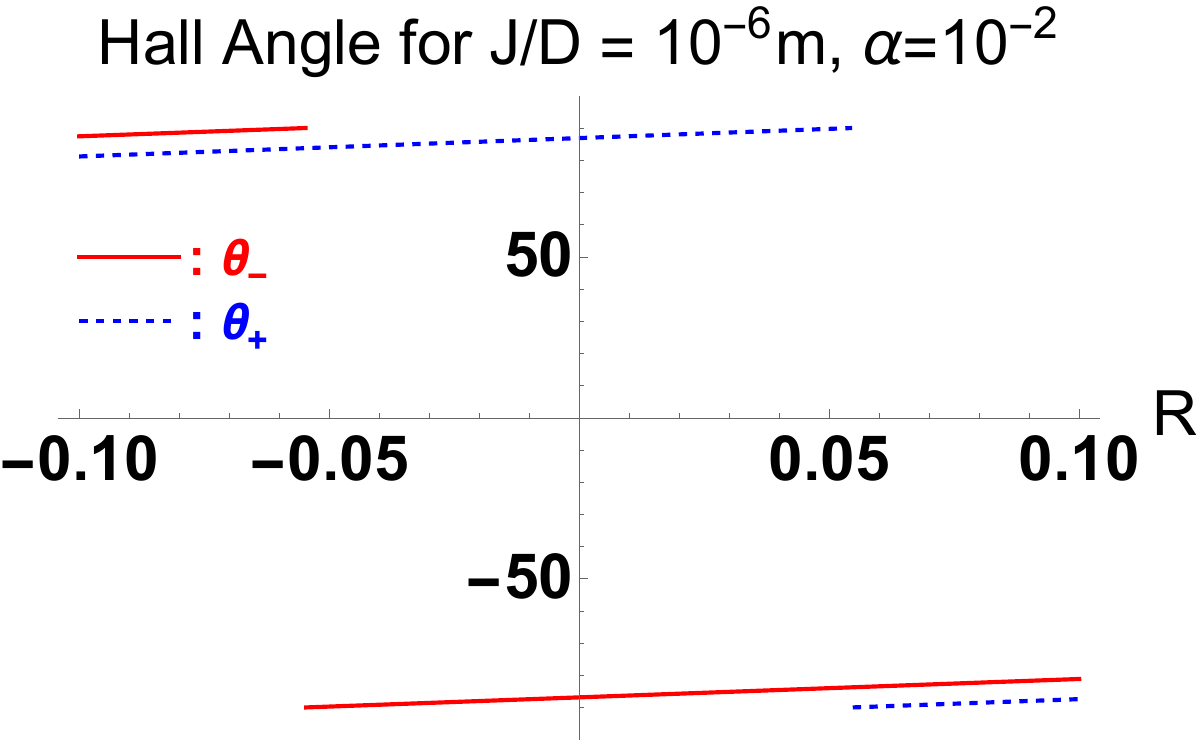}
		\caption{\footnotesize\small  Hall angles $ \theta_\pm (R)$ for $ -0.1 < R <0.1$ \vspace{-0.2in} 
		}
		\label{fig:HallSkyrmionCampared}
	\end{center}
\end{figure}

The generalized Hall angle, in terms of $J/D, \alpha, P$, is 
\begin{align} \label{HallAngleDetailWithR}
	\tan \theta_{\pm} = \frac{[\pm 4 + \alpha C_\alpha R] C_A(J/D) + \pi [ \mp \alpha C_\alpha + 4R] P  }{ [\alpha C_\alpha \mp 4 R]  C_A (J/D)+ \pi [4 \pm  \alpha C_\alpha R] P }   \;. 
\end{align} 
Previous estimates for $R$ is order of $10^{-2}$ \cite{Kim:2020piv}. Moreover, $R$ in \eqref{SkyrmionCM} is dimensionless and is expected to be small. To see its effect, we study the Hall angle for the range, $ -0.1 < R <0.1 $. 
The result \eqref{HallAngleDetailWithR} shows that the parameter $R$ only mixes with the combination of number $4 $ and $\alpha C_\alpha $. Thus the effects of $R$ enhances when $\alpha$ increases. This can be checked in the top two plots in Fig. \ref{fig:HallSkyrmionCampared}, where the two Hall angles $\theta_\pm$ almost coincide in the top left inset with $\alpha=10^{-5}$ compared to the right one with $\alpha=10^{-2}$. 
 
We also notice that there can be a jump of the Hall angle from $\theta = 90^o$ to $- 90^o $ when $ \alpha C_\alpha < 4R$ and the denominator of \eqref{HallAngleDetailWithR} flips sign passing through $0$. It is surprising that this can happen with a relatively small value of $\alpha$. This can be checked in the bottom plots in Fig. \ref{fig:HallSkyrmionCampared}. In the bottom left inset with $J/D = 10^{-6} m, \alpha=10^{-5} $, this jump happens for $\alpha = -0.037$ for $\theta_+$ while there is no jump for $\theta_-$ for the studied range. In the bottom right inset with $J/D = 10^{-6} m, \alpha=10^{-2} $, this jumps happen for $\alpha = -0.055$ for $\theta_-$ and for $\alpha = 0.055$ for $\theta_+$.

\section{Outlook} 

A Hall viscosity term, transverse to the driving force, e.g. temperature gradient, is identified for the skyrmion motion. It depends on the strength of DM interaction and size of the skyrmion, but is independent of skyrmion velocity. It is nothing but the parity-symmetry breaking DM interaction term, after time averaging the circular symmetric magnon contribution. In real materials with temperature gradient, megnons cannot be perfectly circular symmetric. Thus in practice, the strength of the Hall viscosity term is expected to be bigger than our estimates. In this sense, this Hall viscosity term is universal and plays role as far as DM interaction and magnon contributions are involved. 

We showed that this DM Hall viscosity term depends on the skyrmion charge. Thus all the transverse terms, Magnus force and the viscosity, in the Thiele equation depend on the skyrmion charge. It is expected as the two parity breaking sources, magnetic field and DM term, are related to magnetic properties. The Hall viscosity related to this charged dynamics is quite interesting. On the other hand, it is also tempting to look into the possibility for the Hall viscosity effect for neutral dynamics. Is there neutral Hall viscosity effect when the parity symmetry is broken by magnetic field and DM interaction? This is an interesting question along with much recent investigation related to charge neutral Hall viscosity.    

We introduced the neutral Hall viscosity effect, that is proportional to the magnitude of skyrmion velocity (with transverse direction), by the parameter $R$. We propose that the neutral Hall viscosity is independent of the skyrmion charge and thus provides an asymmetry for the skyrmon Hall effect. We checked that the bigger the Gilbert damping parameter $\alpha$, the bigger the neutral Hall viscosity effect. For the parameter range where the skyrmion Hall is larger $\sim 80$, there can be a surprising signature for the Hall angle: the Hall angles can jump from $\theta = 90^o$ to $\theta = -90^o$ as $R$ varies. This can be also achieved by changing other parameters including $J/D$ and $ \alpha$, which is clear in \eqref{HallAngleDetailWithR}. It will be interesting to verify this behavior experimentally.   
\\

\noindent {\it Acknowledgments:} 
I am grateful to Se Kwon Kim and Ren Cheng for useful discussions related to Hall viscosity in the magnetic skyrmion systems. This work has been partially supported by Wisys Spark grant and UW-Parkside summer research funds.


\onecolumngrid 

\indent 
\appendix 

\section{Evaluating LLG equation}    \label{app:EvaluatingLLG} 

Let us compute the second term in the LLG equation \eqref{LLGeq}. The variation of Hamiltonian with a vector $\vec m$ cam be done in terms of index notation. For example, the DM interaction term in Hamiltonian can be evaluated as 
\begin{align}  
	H_{\text{eff},i} = - \frac{\delta (D \epsilon_{jlm} m_j \nabla _l m_m)}{\delta m_i} =- D \epsilon_{ilm} \nabla _l m_m + D \epsilon_{jli} \nabla _l  m_j = -2 D \epsilon_{ilm} \nabla _l m_m \;,
\end{align}	 
up to a total derivative term. Here the index $i$ is a vector index, and we used the epsilon tensor for the cross product $(\vec \nabla_\alpha \times  \vec m )_i= \epsilon_{ilm} \nabla _l m_m $. We write the expansion of the effective field $\vec H_{\text{eff}}$
\begin{align}
\vec H_{\text{eff}} &= J ( \nabla_\alpha^2 \vec m)  -  2D  \vec \nabla_\alpha \times  \vec m + \vec H \nonumber \\
&=  J \Big\{  (1 - m_f^2)^{1/2}  \nabla_\alpha^2 \vec m_s - \frac{(m^i_f \nabla_\alpha  m^i_f)}{ (1 - m_f^2)^{1/2} } \nabla_\alpha  \vec m_s 
- \nabla_\alpha (\frac{(m^i_f \nabla_\alpha  m^i_f)}{ (1 - m_f^2)^{1/2} } )  \vec m_s  +  \nabla_\alpha^2 \vec m_f 
\Big\}  \nonumber \\ 
&~~-  2D \Big\{  (1 - m_f^2)^{1/2} \vec \nabla_\alpha \times  \vec m_s  + \vec \nabla_\alpha \times  \vec m_f  
- \frac{(m^i_f \vec \nabla_\alpha  m^i_f)}{ (1 - m_f^2)^{1/2} } \times  \vec m_s  \Big\}+ \vec H \;. 
\end{align}
The term $\vec m \times \vec H_{\text{eff}}$ after the temporal averaging becomes  
\begin{align}  \label{mHComputation0}
&\vec m \times  \vec H_{\text{eff}} \approx  (1 - \langle m_f^2 \rangle_T)^{1/2}  \vec m_s \times \vec H   \nonumber \\
&+  J \Big\{ (1 - \langle m_f^2 \rangle_T) ~ \vec m_s \times \nabla^2 \vec m_s   
-(1/2) (\nabla_\alpha \langle m^2_f\rangle_T) ( \vec m_s \times \nabla_\alpha  \vec m_s)   +  \langle \vec m_f \times \nabla_\alpha^2 \vec m_f \rangle_T  \Big\}  \nonumber \\ 
&~~ -  2D \Big\{ (1 - \langle m_f^2 \rangle_T) ~  \vec m_s \times (\vec \nabla \times  \vec m_s)   
- (1/2) \vec m_s \times (\vec \nabla \langle m_f^2 \rangle_T \times  \vec m_s)  + \langle  \vec m_f \times (\vec \nabla \times  \vec m_f ) \rangle_T  \Big\}  \;. 
\end{align} 
Here several terms involved with two derivatives on $\vec m_f$ drop out as $ \vec m_s \times  \vec m_s =0 $, and the terms with odd powers of $\vec m_f$ also drop out with temporal averaging. We note that the result is valid for all orders of $\vec m_f$ after the temporal averaging with the magnon profile \eqref{MagnonWaveFunctionCircular}. We also note that the term proportional to $(-1/2)( \nabla_\alpha \langle  \vec m_f^2 \rangle_T ) (\vec m_s \times \nabla_\alpha \vec m_s)$ in \eqref{mHComputation0} is canceled out by a contribution that is from the other term $ \langle \vec m_f \times \nabla_\alpha^2 \vec m_f \rangle_T  $ for circular symmetric magnons. Of course, these two terms cannot be canceled out for realistic situation such as the magnos under temperature gradient. 

For the rest of this section, we look into this by introducing some magnon currents that are discussed in literature, e.g., \cite{MagnonCourseGraining1,MagnonCourseGraining2,MagnonCourseGraining3,MagnonCourseGraining6}. The last term in the second line of \eqref{mHComputation0} can be rewritten as $J ~\langle \vec m_f \times \nabla_\alpha^2 \vec m_f \rangle_T  = J ~\nabla_\alpha \vec  j^f_\alpha  $, where $\vec  j^f_\alpha =  \vec m_f \times \nabla_\alpha \vec m_f$ is perpendicular to $\vec m_f$ and is a tensor with spin index and also space index $\alpha$. This spin current $\vec  j^f_\alpha =  \vec m_f \times \nabla_\alpha \vec m_f$ can be decomposed into components parallel and perpendicular to $\vec m_s$ as $\vec  j^f_\alpha  = \vec m_s [\vec  j^f_\alpha \cdot \vec m_s ] +  (\vec m_s \times \vec m_f) [\vec  j^f_\alpha \cdot (\vec m_s \times \vec m_f)]/m_f^2 $. After a little of algebra, we get  
\begin{align} \label{MagnonCurrentDecomposition}
	\vec m_f \times \nabla_\alpha \vec m_f 
	&= \vec m_s [(\vec m_f \times \nabla_\alpha \vec m_f)\cdot \vec m_s ]  
	+  \vec m_s \times [ \vec m_f (\nabla_\alpha \vec m_s \cdot \vec m_f)] \;.
\end{align}
We also use $\nabla_\alpha (\vec m_s \cdot \vec m_f) = \nabla_\alpha \vec m_s \cdot \vec m_f + \nabla_\alpha \vec m_f \cdot \vec m_s =0$. Note that one cannot discard $(\nabla_\alpha \vec m_s) \cdot \vec m_f$ while keeping $(\nabla_\alpha \vec m_f) \cdot \vec m_s $ by arguing that $\vec m_s$ is slowing varying compared to the fast mode $\vec m_f$ as discussed in the introduction. 	

We evaluate the second term in \eqref{MagnonCurrentDecomposition} for the circular magnons with the wave function \eqref{MagnonWaveFunctionCircular}. We use $\vec \nabla \vec m_s = [\hat \rho \partial_\rho \theta  + (\hat \varphi/\rho) \partial_\varphi \theta ] ~\hat e_1 + [\hat \rho \partial_\rho \phi  + (\hat \varphi/\rho) \partial_\varphi \phi ] \sin \theta ~\hat e_2 = \hat \rho \partial_\rho \theta  ~\hat e_1 + (\hat \varphi/\rho) \partial_\varphi \phi \sin \theta ~\hat e_2 $. Then, 
\begin{align} \label{TemporalAverageMagnonCurrent}
	\langle \vec m_f (\vec \nabla_\alpha \vec m_s  \cdot \vec m_f)  \rangle_T  &= \langle \psi_1^2 \rangle_T \hat \rho \partial_\rho \theta ~\hat e_1 +  \langle \psi_2^2 \rangle_T  (\hat \varphi/\rho) \partial_\varphi \phi \sin \theta ~\hat e_2 \;, \nonumber \\
	&=  \frac{1}{2} \langle \vec m_f^2 \rangle_T ~\vec \nabla_\alpha  \vec m_s \;, 
\end{align}
where we use $\langle \psi_1^2 \rangle_T  = \langle \psi_2^2 \rangle_T  = \langle \psi^2 \rangle_T/2 =  \langle \vec m_f^2 \rangle_T/2$. In the right hand side of the first line, $\partial_\rho$ and $ \partial_\varphi$ are parts of the vector $\vec \nabla_\alpha$. Then, $\langle \vec m_f \times \nabla_\alpha \vec m_f \rangle_T 
= \vec m_s (\langle \vec m_s  \cdot (\vec m_f \times \nabla_\alpha \vec m_f) \rangle_T)  +  ( \langle \vec m_f^2 \rangle_T /2) ~ (\vec m_s \times ~ \vec \nabla \vec m_s ) $.  
Thus, 
\begin{align} \label{MagnonCurrentDecompositionII}
	J ~\langle \vec m_f \times \nabla_\alpha^2 \vec m_f \rangle_T  &= J ~\nabla_\alpha \langle \vec m_f \times \nabla_\alpha \vec m_f \rangle_T \nonumber \\  
	&=J~ j_\alpha (\nabla_\alpha \vec m_s)   
	+ J~ \langle \vec m_f \cdot \nabla_\alpha \vec m_f \rangle_T  ~ (\vec m_s \times \vec \nabla_\alpha  \vec m_s )  
	+ J~ (\langle \vec m_f^2 \rangle_T/2)  ~ (\vec m_s \times \vec \nabla_\alpha^2  \vec m_s )  \;, 
\end{align}
where we use $ \nabla_\alpha j_\alpha =0 $ and define the magnon current as 
\begin{align}
	j_\alpha = \langle \vec m_s  \cdot (\vec m_f \times \nabla_\alpha \vec m_f) \rangle_T \;. \qquad 
\end{align}
The last term in \eqref{MagnonCurrentDecompositionII} can be added to $\vec H_{\text{eff}}^{0s}$ that is defined below \eqref{LLGeqWithMagnonContributions}. The LLG equation for the slow mode reads 
\begin{align} \label{LLGeqWithNotSymmetricMagnonContributionsUpdated}
	\dot {\vec m}_s &= -\gamma \vec m_s \times \vec H_{\text{eff}}^{s} +  (1 - \langle m_f^2 \rangle)^{1/2} \cdot \alpha \vec m_s \times \dot {\vec m}_s  \\
	& - \frac{\gamma }{(1 - \langle m_f^2 \rangle)^{1/2} } \Big\{ - J~ \langle \vec m_f \cdot \nabla_\alpha \vec m_f \rangle_T  ~ (\vec m_s \times \vec \nabla_\alpha  \vec m_s )  + J~ j_\alpha (\nabla_\alpha \vec m_s) + J~ \langle \vec m_f \cdot \nabla_\alpha \vec m_f \rangle_T  ~ (\vec m_s \times \vec \nabla_\alpha  \vec m_s )    \nonumber  \\ &\qquad\qquad\qquad\qquad 
	- 2D ~ \langle \vec m_f \times ( \nabla \times \vec m_f) \rangle_T 
	+ 2D ~\vec m_s \times [( \vec \nabla \langle m_f^2 \rangle_T) \times \vec m_s   \Big\}
	\;. \nonumber 
\end{align} 	 
Here, $\vec H_{\text{eff}}^{s}$ contains the terms with derivatives only on the slow mode, and slightly modified as $ \vec H_{\text{eff}}^{s} =  \vec H + J  \frac{1 - \langle m_f^2 \rangle/ 2}{(1 - \langle m_f^2 \rangle)^{1/2}}  ~(\nabla^2 \vec m_s ) -  2D (1 - \langle m_f^2 \rangle)^{1/2} (\vec \nabla \times  \vec m_s) $. Note that the first and last terms in the second line cancel each other. Thus the Thiele equation simplifies for the circular symmetric magnons. For more realistic magnons in the presence of temperature gradient, the magnon profile cannot be completely circular symmetric as it transfers energy from the hot region to the cold region. This means $\langle \psi_1^2 \rangle_T \neq \langle \psi_2^2 \rangle_T $, and thus \eqref{TemporalAverageMagnonCurrent} does not holds.

\section{Evaluating Thiele equation}   \label{app:ExampleThieleEq} 

In this appendix, we show explicit computations for contracting $(\vec m_s \times \nabla_\beta \vec m_s) \cdot $ on the LLG equation \eqref{LLGeqWithMagnonContributions}, with a dot product acting on the spin vector index, to get  \eqref{ThieleEqMagnonAveraged}, followed by the computation leading to \eqref{ThieleEqMagnonCoordinates} using the local coordinates listed in \eqref{LocalCoordinatesI} and \eqref{MagnonWaveFunctionCircular}. We write again the LLG equation \eqref{LLGeqWithMagnonContributions} for the slow mode so that the appendix can be read in self-contained manner.  
\begin{align} \label{LLGeqWithMagnonContributionsApp}
	0 =& - \dot {\vec m}_s +  (1 - \langle m_f^2 \rangle)^{1/2} \cdot \alpha \vec m_s \times \dot {\vec m}_s    \\
	& -\gamma \vec m_s \times \Big\{  \vec H + J (1 - \langle m_f^2 \rangle)^{1/2} ~(\nabla^2 \vec m_s  ) 
	-  2D (1 - \langle m_f^2 \rangle)^{1/2} (\vec \nabla \times  \vec m_s)  \Big\}  \nonumber \\
	& - \frac{\gamma }{(1 - \langle m_f^2 \rangle)^{1/2} } \Big\{  J ~\langle \vec m_f \times \nabla_\alpha^2 \vec m_f \rangle_T  
	- J ~ \langle m_f^i \nabla_\alpha m_f^i \rangle_T [\vec m_s \times  \nabla_\alpha \vec m_s]    \nonumber  \\
	&\qquad\qquad\qquad\qquad - 2D ~ \langle \vec m_f \times ( \nabla \times \vec m_f) \rangle_T 
	+ 2D ~\vec m_s \times [( \vec \nabla \langle m_f^2 \rangle_T) \times \vec m_s   \Big\}
	\;.  \nonumber
\end{align} 	 
The first line of \eqref{LLGeqWithMagnonContributionsApp} contains the terms with $\dot {\vec m}_s$, a time derivative  of the slow mode. Then 
\begin{align}
	(\vec m_s \times \nabla_\alpha \vec m_s) \cdot \dot {\vec m}_s &= \epsilon_{ijk} ( m_s^j \nabla_\alpha m_s^k ) \dot { m}_s^i = {\vec m}_s \cdot  (\nabla_\alpha \vec m_s \times \nabla_\beta \vec m_s) v_\beta \equiv  \mathcal G_{\alpha\beta} v_\beta \;,  
\end{align}
where $\epsilon_{ijk} $ is the totally antisymmetric tensor with $\epsilon_{123} =1$. $ \dot {\vec m}_s = v_\alpha \nabla_\alpha \vec m_s$ with a space index $\alpha, \beta = x,y$ or $ 1,2$.  
\begin{align}
&(\vec m_s \times \nabla_\alpha \vec m_s) \cdot (\vec m_s \times \dot {\vec m}_s ) = \epsilon_{ijk} \epsilon_{ilm}  ( m_s^j \nabla_\alpha m_s^k ) ( m_s^l \nabla_\beta m_s^m )~ v_\beta \nonumber \\
&\qquad = {m}_s^2 ~  (\nabla_\alpha \vec m_s \cdot \nabla_\beta \vec m_s) v_\beta -  (\vec {m}_s \cdot \nabla_\alpha \vec m_s) ( \vec m_s \cdot \nabla_\beta \vec m_s) v_\beta 
= (\nabla_\alpha \vec m_s \cdot \nabla_\beta \vec m_s) ~v_\beta  \equiv  \mathcal D_{\alpha\beta} v_\beta \;,  \nonumber 
\end{align}
where we use $ m_s^2 = 1$, $ \vec {m}_s \cdot \nabla_\alpha \vec m_s = \nabla_\alpha m_s^2 /2= 0$, and the identity $\epsilon_{ijk} \epsilon_{ilm} = \delta_{jl} \delta_{km} - \delta_{jm} \delta_{kl}$. These two terms are in the first line of \eqref{ThieleEqMagnonAveraged}. The second line of \eqref{LLGeqWithMagnonContributionsApp} contains the usual terms without the magnon contribution with modification with the magnitude of magnons $ (1 - \langle m_f^2 \rangle)^{1/2} $. Using $ (\vec m_s \times \nabla_\alpha \vec m_s) \cdot (\vec m_s \times \vec A ) =(\vec A \cdot \nabla_\alpha m_s ) - (\vec m_s \cdot \vec A)( \vec m_s \cdot \nabla_\alpha \vec m_s  ) =(\vec A \cdot \nabla_\alpha m_s )$, we arrive the second line of \eqref{ThieleEqMagnonAveraged}. 

At this point, we evaluate the first and second lines of \eqref{ThieleEqMagnonAveraged} by integrating over a unit skyrmion size. We use the local coordinates \eqref{LocalCoordinatesI}, \eqref{MagnonWaveFunctionCircular} and the skyrmion profile \eqref{SmoothSkyrmionModel}. $\mathcal G_{\alpha\beta}$ is anti-symmetric ($\mathcal G_{xx} = \mathcal G_{yy}=0$) and we only evaluate $\mathcal G_{xy}$.   
\begin{align}
	\int dV \mathcal G_{xy} v_y 
	&= \int_0^{2P} \rho d\rho \int_0^{2\pi} d\varphi  ~ \hat e_3 \cdot (\nabla_x \hat e_3 \times \nabla_y \hat e_3) v_y
	= \int_0^{2P} \rho d\rho \int_0^{2\pi} d\varphi  ~ \frac{1}{\rho} (\partial_\rho \theta (\rho))(\partial_\varphi \phi (\varphi)) \sin \theta(\rho) v_y \nonumber \\
	&= -2\pi  \int_0^{2P} d\rho (\partial_\rho \cos \theta (\rho) v_y = -2\pi \cos \theta (\rho) \Big|_{0}^{2P} v_y= 4\pi v_y \;, 
\end{align}
where we use 
\begin{align}
	\nabla_x \hat e_3 &= (\partial_x \rho) (\partial_\rho \theta) \hat e_1 +  (\partial_x \varphi) (\partial_\varphi \phi) \sin \theta \hat e_2 \;, \nonumber \\
	\nabla_y \hat e_3 &= (\partial_y \rho) (\partial_\rho \theta) \hat e_1 +  (\partial_y \varphi) (\partial_\varphi \phi) \sin \theta \hat e_2 \;, \nonumber \\ 
	 \mathcal G_{xy} &= \hat e_3 \cdot (\nabla_x \hat e_3 \times \nabla_y \hat e_3) = \{ (\partial_x \rho) (\partial_y \varphi) - (\partial_x \varphi) (\partial_y \rho)  \}  (\partial_\rho \theta) (\partial_\varphi \phi) \sin \theta = \frac{1}{\rho}  (\partial_\rho \theta) (\partial_\varphi \phi) \sin \theta \;, \nonumber 
\end{align}
and $ \theta = \theta (\rho), \phi(\varphi) = \varphi  $, $x = \rho \cos \varphi, y = \rho \sin \varphi  $.

The symmetric term with $ \mathcal D_{\alpha\beta} =\nabla_\alpha \vec m_s \cdot \nabla_\beta \vec m_s$ can be evaluated similarly. For example the term with $ \mathcal D_{xx} $ is   
\begin{align} 
&\alpha \int dV (1 - \langle m_f^2 \rangle)^{1/2} (\nabla_x \vec m_s \cdot \nabla_x \vec m_s) ~v_x   \nonumber \\
&= \alpha v_x~(1 - \langle \psi^2 \rangle)^{1/2} \int_0^{2P} \rho d\rho \int_0^{2\pi} d\varphi \left\{ \cos^2 \varphi (\partial_\rho \theta)^2 + \sin^2 \varphi \frac{\sin^2 \theta}{\rho^2}  \right\}  \nonumber \\
&= \alpha v_x~(1 - \langle \psi^2 \rangle)^{1/2} \times \pi \left\{ \int_{P-{\omega_{\text{DW}}}/2}^{P+{\omega_{\text{DW}}}/2} \rho d\rho \left( \frac{\pi}{{\omega_{\text{DW}}} }\right)^2 + \int_{P-{\omega_{\text{DW}}}/2}^{P+{\omega_{\text{DW}}}/2} d\rho  \frac{\sin^2 \theta}{\rho}  
 \right\}   \nonumber \\
&= \alpha v_x~(1 - \langle \psi^2 \rangle)^{1/2} \times \pi \left\{  \frac{\pi^2 P}{{\omega_{\text{DW}}}} +C_{\omega_{\text{DW}}} \right\} 
\;, 
\end{align}
where we assume that the magnon profile does not change much over the unit skyrmion size, so the factor $ (1 - \langle \psi^2 \rangle)^{1/2}$ can come out of the integral. We also use $\partial_\rho \hat e_3 = \partial_\rho\theta (\rho) \hat e_1 $ and $\partial_\varphi \hat e_3 =  \sin\theta(\rho) \hat e_2 $ as $\partial_\varphi \phi = 1$ with the identification $\phi = \varphi$. Then, $\mathcal D_{xx} = \nabla_x \hat e_3 \cdot \nabla_x \hat e_3 = ((\nabla_x \rho) (\partial_\rho \theta)  ~\hat e_1 + (\nabla_x \varphi) (\partial_\varphi \phi) \sin \theta ~\hat e_2 )^2 = \cos^2 \varphi (\partial_\rho \theta)^2 + \frac{\sin^2 \varphi}{\rho^2} \sin^2 \theta $. 
\begin{align} \label{CDIntegral}
C_{{\omega_{\text{DW}}}} &= \int_{P-{\omega_{\text{DW}}}/2}^{P+{\omega_{\text{DW}}}/2} d \rho  \frac{\sin^2 \theta}{\rho} =\frac{1}{2 } \log \frac{(2 P +{\omega_{\text{DW}}} )}{(2 P-{\omega_{\text{DW}}} )}  
+\frac{1}{2 }  \cos \big( \frac{2 \pi  P}{{\omega_{\text{DW}}}} \big) \{ \text{Ci}\left(\frac{2 \pi  P}{{\omega_{\text{DW}}}} +\pi \right) - \text{Ci}\left( \frac{2 \pi  P}{D}- \pi \right)\}  \nonumber \\
&\hspace{1.5in}+ \frac{1}{2 }  \sin \big(\frac{2 \pi  P}{{\omega_{\text{DW}}}} \big) \{ \text{Si} \left(\frac{2 \pi  P}{{\omega_{\text{DW}}}} +\pi \right)-\text{Si} \left( \frac{2 \pi  P}{{\omega_{\text{DW}}}} -\pi \right) \} \;. \nonumber 
\end{align}
Here the integral range is reduced to the domain wall inside the skyrmion as $ \sin 0 =\sin \pi=0$. This $C_{{\omega_{\text{DW}}}}$ is typically much smaller compared to the first term $\pi^2 P/{\omega_{\text{DW}}}$. It depends only on the combination $ P/{\omega_{\text{DW}}}$ and decreases as the skyrmion DW region gets skinnier (meaning increasing $ P/{\omega_{\text{DW}}}$).

The three terms in the second line of \eqref{ThieleEqMagnonAveraged} vanish when evaluating with the local coordinates for $\alpha = x$ or $y$. 
\begin{align} 
\gamma \int dV ~(\nabla_x \vec m_s) \cdot \vec H  &=\gamma \int dV ~\big( (\partial_x \rho) (\partial_\rho \theta) \hat e_1 +  (\partial_x \varphi) (\partial_\varphi \phi) \sin \theta \hat e_2 \big)   \cdot \hat z H \nonumber \\
&=- \gamma H \int dV ~ (\partial_x \rho) (\partial_\rho \theta) \sin \theta  \propto \int_0^{2\pi} d\varphi \sin \varphi  = 0 \;, \\
-\gamma J (1-m_f^2)^{1/2}  \int dV (\nabla_x \vec m_s \cdot \nabla^2 \vec m_s) &= \int dV ~ (\partial_x \rho) (\cdots) = (\text{terms independent of $\varphi$} ) \times \int_0^{2\pi} d\varphi \sin \varphi   = 0 \;, 
\\
2D \gamma (1-m_f^2)^{1/2}  \int dV  (\nabla_x \vec m_s \cdot \vec \nabla \times \vec m_s) &= \int dV ~ (\partial_x \varphi) (\cdots) = (\text{terms independent of $\varphi$} ) \times \int_0^{2\pi} d\varphi \sin \varphi  = 0 
\;. 
\end{align}
Here the first term is evaluated for constant magnetic field along $\hat z$ direction $ \vec H = \hat z H$ along with $\hat e_1 \cdot \hat z = \sin \theta, \partial_x \rho = \cos \varphi $. The second term also vanishes similarly as 
$ \nabla^2 \vec m_s = \{ (\partial_\rho \theta)/\rho  + \partial_\rho^2 \theta - \sin\theta \cos \theta /\rho^2 \} \hat e_1- \{ (\partial_\rho \theta)^2 + \sin^2\theta /\rho^2 \} \hat e_3$, and thus $\nabla_x \vec m_s \cdot \nabla^2 \vec m_s = (\partial_x \rho) (\partial_\rho \theta) \{ (\partial_\rho \theta)/\rho + \partial_\rho^2 \theta - \sin\theta \cos \theta /\rho^2 \} = \cos\varphi \times (\text{terms independent of $\varphi$})$. 
The third term is evaluated using 
\begin{align}
	\vec \nabla \times \vec m_s &= (\hat \rho \partial_\rho + \frac{\hat \varphi}{\rho} \partial_\varphi ) \times \hat e_3 =
	(\partial_\rho \theta) \hat \rho \times \hat e_1 + \frac{\sin\theta (\partial_\varphi \phi)}{\rho} \hat \varphi \times \hat e_2 \nonumber \\
	&= (\partial_\rho \theta) (\cos\theta \hat e_1 + \sin\theta \hat e_3) \times \hat e_1 + \frac{\sin\theta (\partial_\varphi \phi)}{\rho} \hat e_2 \times \hat e_2 = (\partial_\rho \theta) \sin\theta~ \hat e_2 \;,
\end{align} 
and thus $(\nabla_x \vec m_s \cdot \vec \nabla \times \vec m_s) = (\partial_x \varphi) (\partial_\varphi \phi) (\partial_\rho \theta) \sin^2 \theta  = (\text{terms independent of $\varphi$} ) \times (- \sin \varphi) $. Here we used $ d\varphi = (\cos\varphi dy - \sin \varphi dx)/\rho$ by taking a derivative on both sides of $ \tan \varphi = y/x$. 

Until this point, we show that the first two lines of the Thiele equation \eqref{ThieleEqMagnonAveraged} collapsed into the first line of \eqref{ThieleEqMagnonCoordinates}. We turn to consider the terms that are proportional to $J$ in \eqref{LLGeqWithMagnonContributions}.   

Next we study the term, $\frac{\gamma (J)}{(1 - \langle m_f^2 \rangle)^{1/2} } ~ \langle m_f^i \nabla_\alpha m_f^i \rangle_T [\vec m_s \times  \nabla_\alpha \vec m_s] $, in LLG equation \eqref{LLGeqWithMagnonContributions} to arrive the corresponding term in \eqref{ThieleEqMagnonCoordinates}. By contracting $(\vec m_s \times \nabla_\beta \vec m_s)$ with a dot product acting on the spin vector $[\vec m_s \times  \nabla_\alpha \vec m_s]$ 
\begin{align}
(\vec m_s \times \nabla_\beta \vec m_s) \cdot [\vec m_s \times  \nabla_\alpha \vec m_s]  
= (\nabla_\beta  \vec m_s \cdot \nabla_\alpha \vec m_s)  - (\vec m_s \cdot \nabla_\beta \vec m_s) (\vec m_s \cdot  \nabla_\alpha \vec m_s] 
\;. \nonumber
\end{align} 	 
The last term vanishes as $ \nabla_\alpha m_s^2 = 0$ and thus $ \vec m_s \cdot \nabla_\alpha \vec m_s = 0 $. 

Now we integrate the term over a unit volume of the skyrmion $\gamma (J) \int dV  \frac{ (\vec m_f \cdot \nabla_\beta \vec m_f) (\nabla_\beta  \vec m_s \cdot \nabla_\alpha \vec m_s)   }{(1 - \langle m_f^2 \rangle)^{1/2} } $ with $\mathcal D_{\beta\alpha} = \nabla_\beta  \vec m_s \cdot \nabla_\alpha \vec m_s$. We consider the magnon profile has non-zero contribution along $x$ direction, for example with temperature gradient along $x$ direction, so that $ \nabla_x \langle \psi^2 \rangle \neq 0$, while $ \nabla_y \langle \psi^2 \rangle = 0$ in the 2 dimensional plane with coordinates $(x,y)$. Then 
$$
\langle \vec m_f \cdot \nabla_\beta \vec m_f \rangle_T= (1/2) \nabla_x \langle \psi^2 \rangle \;. 
$$
We also can check that $ \mathcal D_{\alpha\beta} \propto \delta_{\alpha\beta}$. And 
\begin{align}
\mathcal D_{xx} &= \nabla_x \hat e_3 \cdot \nabla_x \hat e_3 = ((\nabla_x \rho) (\partial_\rho \theta)  ~\hat e_1 + (\nabla_x \varphi) (\partial_\varphi \phi) \sin \theta ~\hat e_2 )^2  \nonumber \\
&= \cos^2 \varphi (\partial_\rho \theta)^2 + \frac{\sin^2 \varphi}{\rho^2} \sin^2 \theta  \;,
\end{align}
where we use $\partial_\rho \hat e_3 = \partial_\rho\theta (\rho) \hat e_1 $ and $\partial_\varphi \hat e_3 =  \sin\theta(\rho) \hat e_2 $ as $\partial_\varphi \phi = 1$ with the identification $\phi = \varphi$. 
\begin{align} 
&\gamma (J) \int dV  \frac{ (\vec m_f \cdot \nabla_\beta \vec m_f) (\nabla_\beta  \vec m_s \cdot \nabla_\alpha \vec m_s)   }{(1 - \langle m_f^2 \rangle)^{1/2} }  \nonumber \\
&= \hat x~ \gamma (J) ~ \frac{  \nabla_x \langle \psi^2 \rangle / 2}{(1 - \langle \psi^2 \rangle)^{1/2} } 
\int_0^{2P} \rho d\rho \int_0^{2\pi} d\varphi \left\{ \cos^2 \varphi (\partial_\rho \theta)^2 + \frac{\sin^2 \varphi}{\rho^2} \sin^2 \theta \right\}  \nonumber \\
&= \hat x~ \gamma (J) ~ \frac{  \nabla_x \langle \psi^2 \rangle}{(1 - \langle \psi^2 \rangle)^{1/2} } \times 
\frac{\pi}{2} \left\{ \pi^2 \frac{P}{{\omega_{\text{DW}}}} +C_{\omega_{\text{DW}}} \right\} 
\;, \nonumber
\end{align}
where 
\begin{align} \label{CDIntegral2}
C_{{\omega_{\text{DW}}}} &= \int_{P-{\omega_{\text{DW}}}/2}^{P+{\omega_{\text{DW}}}/2} d \rho  \frac{\sin^2 \theta}{\rho} =\frac{1}{2 } \log \frac{(2 P +{\omega_{\text{DW}}} )}{(2 P-{\omega_{\text{DW}}} )}  
+\frac{1}{2 }  \cos \big( \frac{2 \pi  P}{{\omega_{\text{DW}}}} \big) \{ \text{Ci}\left(\frac{2 \pi  P}{{\omega_{\text{DW}}}} +\pi \right) - \text{Ci}\left( \frac{2 \pi  P}{D}- \pi \right)\}  \nonumber \\
&\hspace{1.5in}+ \frac{1}{2 }  \sin \big(\frac{2 \pi  P}{{\omega_{\text{DW}}}} \big) \{ \text{Si} \left(\frac{2 \pi  P}{{\omega_{\text{DW}}}} +\pi \right)-\text{Si} \left( \frac{2 \pi  P}{{\omega_{\text{DW}}}} -\pi \right) \} \;. \nonumber 
\end{align}
Here we evaluate the slow and fast modes independently by assuming $\nabla_x \langle \psi^2 \rangle_T $ is constant over the unit skyrmion volume. 

Similarly, the other term $\gamma (J) \int dV  \frac{ [ \vec m_s \cdot \nabla^2 \vec m_f] [\vec m_f \cdot  \nabla_\alpha \vec m_s]  }{(1 - \langle m_f^2 \rangle)^{1/2} }$ that contributes to $x$ direction can be evaluated. Using 
\begin{align}
&\vec m_s \cdot \nabla^2 \vec m_f = - \frac{1}{\rho} \Big\{ 2\rho (\partial_\rho \psi_1) (\partial_\rho \theta) - (\partial_\rho \theta) \psi_1 + \rho (\partial_\rho^2 \theta) \Big\} + \frac{1}{\rho^2} \Big\{ (\partial_\varphi \phi) \sin\theta \cos\theta \psi_1 + 2 \sin \theta (\partial_\rho \psi_2) \Big\} \;,  \\
&\vec m_f \cdot  \nabla_\beta \vec m_s = (\delta_{\beta\rho}) \hat \rho (\partial_\rho \theta) \psi_1  + (\delta_{\beta\varphi}) (\hat \varphi/\rho) \sin\theta \psi_2  \;., 
\end{align}
and after taking the temporal coarse graining, we get 
\begin{align}
\langle [ \vec m_s \cdot \nabla^2 \vec m_f] [\vec m_f \cdot  \nabla_\alpha \vec m_s]  \rangle_T &= 
- (\delta_{\alpha\rho}) \hat \rho (\partial_\rho \theta)^2  \Big\{  \partial_\rho \langle \psi_1^2 \rangle_T + \frac{\psi_1^2}{\rho} + \frac{\sin\theta \cos\theta }{\rho^2} \psi_1^2
\Big\} - (\delta_{\alpha\varphi}) (\hat \varphi) \frac{\sin\theta }{\rho} \partial_\varphi \langle \psi_2^2 \rangle_T \nonumber \\
&= - \hat x \Big\{ (\partial_\rho \theta)^2 \nabla_x \langle \psi_1 ^2 \rangle \cos^2 \varphi + \nabla_x \langle \psi_2 ^2 \rangle \frac{\sin^2 \varphi}{\rho^2}  \Big\}  \;.
\end{align}
This can be decomposed into $x$ and $y$ coordinates. By keeping the contributions for $\nabla_x \langle \psi_1^2 \rangle  =\nabla_x \langle \psi_2^2 \rangle = \nabla_x \langle \psi^2 \rangle /2$, we get 
\begin{align}
&\gamma (J) \int dV  \frac{ [ \vec m_s \cdot \nabla^2 \vec m_f] [\vec m_f \cdot  \nabla_\alpha \vec m_s]  }{(1 - \langle m_f^2 \rangle)^{1/2} }  \nonumber \\
&=- \hat x~ \gamma (J) ~ \frac{  \nabla_x \langle \psi^2 \rangle / 2}{(1 - \langle \psi^2 \rangle)^{1/2} } 
\int_0^{2P} \rho d\rho \int_0^{2\pi} d\varphi \left\{ \cos^2 \varphi (\partial_\rho \theta)^2 + \frac{\sin^2 \varphi }{\rho^2}  \right\}  \nonumber \\
&=- \hat x~ \gamma (J) ~ \frac{  \nabla_x \langle \psi^2 \rangle}{(1 - \langle \psi^2 \rangle)^{1/2} } \times 
\frac{\pi}{2} \left\{ \pi^2 \frac{P}{{\omega_{\text{DW}}}} + \log \Big( \frac{ 2P}{\rho_{min}} \Big) \right\} 
\;, \nonumber
\end{align}
Here $4\pi P^2  = \int_0^{2P} \rho d\rho  \int_0^{2\pi} d\varphi  $ is the skyrmion unit volume.

By combining these two contributions, we get 
\begin{align} \label{ThieleEqXPartsApp}
& \hat x ~ \frac{ \gamma (J) ~ \nabla_x \langle \psi^2 \rangle_T}{(1 - \langle \psi^2 \rangle)_T} \Big\{ \frac{\pi}{2} [  \frac{\pi^2 P}{{\omega_{\text{DW}}}} +C_{{\omega_{\text{DW}}}} ]  - \frac{\pi}{2} [\frac{\pi^2  P}{{\omega_{\text{DW}}}} +  \log \Big( \frac{ 2P}{\rho_{min}} \Big) ] \Big\}  \nonumber  \\
&\qquad = - \hat x ~ \frac{ \gamma (J) ~ \nabla_x \langle \psi^2 \rangle_T}{(1 - \langle \psi^2 \rangle)_T} \frac{\pi}{2} \Big\{  \log \Big( \frac{ 2P}{\rho_{min}} \Big) - C_{{\omega_{\text{DW}}}} \Big\} \;. 
\end{align} 
From this analytic result, we see that the longitudinal force on skyrmion due to the magnons depends on the details of skyrmion structure,  in particular the area of domain wall region of the skyrmion, and is opposite direction to the magnon density change.


\begin{thebibliography}{999}
\bibitem{footnote1}
This term contributes to the LLG equation, but drops our of the Thiele equation as $  \langle \vec m_f \times \dot{\vec m}_f \rangle_T  \propto \hat e_3 =  \vec m_s$ using the local coordinates \eqref{LocalCoordinatesI}. Thus, $ (\vec m_s \times \vec \nabla_\beta \vec m_s) \cdot \langle \vec m_f \times \dot{\vec m}_f \rangle_T  \propto (\vec m_s \times \vec \nabla_\beta \vec m_s) \cdot \vec m_s = 0$ in \eqref{ThieleEqMagnonAveraged}.
\bibitem{footnote2}
The terms in \eqref{HallAngleDetail} have the following ranges, $10^{-10} < 4 C_A(J/D) < 10^{-4},  10^{-11} < \pi \alpha C_\alpha P < 10^{-8}, 10^{-15} <  \alpha C_\alpha C_A (J/D) < 10^{-6}$.
\end{thebibliography}
\end{document}